\documentclass{./aa}  
\usepackage{natbib}
\usepackage{graphicx}
\usepackage{txfonts}
\newcommand{\beq}{\begin{equation}}
\newcommand{\eeq}{\end{equation}}
\newcommand{\beqa}{\begin{eqnarray}}
\newcommand{\eeqa}{\end{eqnarray}}

\newcommand{\dw}[1]{_{\scriptstyle \mathrm{#1}}}

\begin{document}
   \title{Flow instabilities of magnetic flux tubes}

   \subtitle{II. Longitudinal flow}

   \author{V. Holzwarth 
          \and
	  D. Schmitt 
	  \and 
	  M. Sch{\"ussler}
}

   \institute{
           Max-Planck-Institut f\"ur Sonnensystemforschung,
           37191 Katlenburg-Lindau, Germany\\
             \email{holzwarth,schmitt,schuessler@mps.mpg.de}
             }
   \date{\today}

\abstract
{Flow-induced instabilities are  relevant for the storage 
 and dynamics of magnetic fields in stellar
 convection zones and possibly also in other astrophysical contexts.}
{We continue the study started
in the first paper of this series by considering the stability 
properties of longitudinal flows along magnetic flux tubes.}
{A linear stability analysis was carried out to 
determine criteria for the onset of instability in the framework of
the approximation of thin magnetic flux tubes.}
{In the non-dissipative case, we find Kelvin-Helmholtz instability for
flow velocities exceeding a critical speed that depends on the
Alfv{\'e}n speed and on the ratio of the internal and external densities. 
Inclusion of a friction term proportional to the relative transverse
velocity leads to a friction-driven instability connected with 
backward (or negative energy) waves. We discuss the physical nature of
this instability. In the case of a stratified external medium, the
Kelvin-Helmholtz instability and the friction-driven 
instability can set in for flow speeds significantly
lower than the Alfv{\'e}n speed.}
{Dissipative effects can excite flow-driven instability below the
thresholds for the Kelvin-Helmholtz and the undulatory (Parker-type) 
instabilities. This may be important for magnetic flux storage in
stellar convection zones and for the stability of astrophysical jets.}

\titlerunning{Flow instabilities of magnetic flux tubes II}
   \maketitle
%

\section{Introduction}
\label{sec:intro}

The stability of magnetic structures with field-aligned flows is of
potential astrophysical relevance in relation to magnetic flux storage
in stellar convection zones, for siphon flows and other flows in stellar
atmospheres and envelopes, as well as for jets and collimated outflows.
While the first paper of this series \citep{Schuessler:Ferriz-Mas:2007}
studied the effects of perpendicular flows on the stability of magnetic
flux tubes, here we consider the case of parallel flows, directed along
the tube%
\footnote{For flows that do not vary along the tube, this covers both
          flows inside and outside the flux tube, since the two cases
	  are related to each other by a simple Galilean transformation
	  of the frame of reference.}.

Linear stability properties of field-aligned flows in ideal MHD have
been considered previously with regard to the Kelvin-Helmholtz
instability by various authors \citep[e.g.,][]{Chandrasekhar:1961,
Parker:1964, Ray:1981, Ferrari:etal:1981, Narayanan:Somasundaram:1982,
Rae:1983, DSilva:Choudhuri:1991, Cheng:1994, Kolesnikov:etal:2004}.
Dissipative effects leading to instability in the presence of
longitudinal (shear) flows, connected with the appearance of negative
energy waves, have been discussed by \citet{Ryutova:1988, Ryutova:1990}
and \citet{Joarder:etal:1997} and, in connection with resonant
absorption of waves in shear layers, by \citet{Hollweg:etal:1990},
\citet{Goossens:etal:1992}, \citet{Tirry:etal:1998}, and
\citet{Andries:Goossens:2001a, Andries:Goossens:2001b}.

The effect of a longitudinal flow along a thin toroidal magnetic flux
tube has been considered by \citet{Ballegooijen:1983} and
\citet{Ferriz-Mas:Schuessler:1993, Ferriz-Mas:Schuessler:1995} in the
case of a rotating star and by \citet{Achterberg:1996b} in the case of a
Keplerian accretion disk. A general formalism for the stability of
stationary flows along curved thin magnetic flux tubes has been
developed by \citet{Schmitt:1998}.

The work presented here is motivated by the unexpected appearance
of instabilities in numerical simulations of thin flux tubes in stellar
convection zones \citep{Holzwarth:2002} in parameter domains where
linear analysis predicts stability
\citep[e.g.,][]{Ferriz-Mas:Schuessler:1995}. The instabilities could be
traced back to the aerodynamic drag force included in the simulations,
which does not appear in linear analysis owing to its quadratic
dependence on the relative velocity between flux tube and
environment. Such instabilities affect the storage of magnetic flux at
the bottom of a stellar convection zone, so that a systematic
investigation of the underlying physical mechanisms and the consequences
for flux storage is called for. Our study is carried out in the
framework of the thin flux tube approximation \citep{Spruit:1981}, which
is relevant for magnetic structures in the deep layers of a stellar
convection zone. This approximation is still the only possibility to
carry out numerical simulations of the dynamics of magnetic structures
in these layers with realistically high values of the plasma $\beta$
(ratio of gas pressure to magnetic pressure), since the correspondingly
low density, temperature, and pressure contrasts are difficult to
maintain against numerical diffusion in 2D/3D MHD simulations.

Of particular importance with respect to the question of flux storage is
the inclusion of stratification effects, so that the combined effects of
flow-driven instability and Parker instability can be studied. We
provide a unified linear treatment of the Kelvin-Helmholtz instability,
dissipative (friction-induced) instability, and the undulatory
(Parker-type) instability in a gravitationally stratified medium.  While
we restrict ourselves here to straight flux tubes in a plane-parallel
stratification in order to elucidate the physical mechanisms, the
subsequent paper in the series will consider the case of toroidal flux
tubes and will quantitatively discuss the relevance of these
instabilities in rotating stars and the dependence on the stratification
parameters, with particular attention to the problem of flux storage in
stellar convection zones.

The paper is organized as follows. We first consider the stability of a
straight flux tube with longitudinal flow in a (non-stratified) uniform
background medium in Sect.~\ref{sec:uniform}. The treatment is extended
in Sect.~\ref{sec:stratification} to include the effect of gravitational
stratification. Section~\ref{sec:conclusion} gives our conclusions.

\section{Uniform background medium}
\label{sec:uniform}

As a first example, we consider a straight, untwisted magnetic flux
tube, which harbors a flow along the field lines with constant velocity,
$u_0$, and is embedded in a uniform and static environment. Under these
circumstances, the equilibrium for a thin flux tube is simply given by
the condition of total pressure balance,
\begin{equation}
   p_{\rm e0} = p_{\rm i0} + {B_0^2\over 8\pi}\,,
   \label{eq:pressbal}
\end{equation}
where $p_{\rm e0}$ and $p_{\rm i0}$ are the gas pressures in the
environment and within the flux tube, respectively, and $B_0={\rm
const.}$ is the magnetic field strength. We indicate equilibrium
quantities by an index ``0''. Following previous work
\citep[e.g.,][]{Spruit:Ballegooijen:1982, Schuessler:1990a,
Ferriz-Mas:Schuessler:1993, Ferriz-Mas:Schuessler:1995, Schmitt:1998}, we
consider a Lagrangian displacement, $\vec{\xi}$, of the equlibrium flux
tube, expressed in terms of the Frenet basis, $\{ \vec{e}_{\rm t0},
\vec{e}_{\rm n0}, \vec{e}_{\rm b0}\} $ in the tangential, normal, and
binormal directions of the equilibrium flux tube,
\begin{equation}$$ \vec{\xi} = \xi_{\rm t} \vec{e}_{\rm t0} + 
         \xi_{\rm n} \vec{e}_{\rm n0} +
         \xi_{\rm b} \vec{e}_{\rm b0} \,.
   \label{eq:perturb}
\end{equation} 
We define the direction of $\vec{e}_{\rm t0}$ such that $u_0 \ge 0$.
For a straight flux tube, $\vec{e}_{\rm n0}$ and $\vec{e}_{\rm b0}$ can
be defined as any pair of orthogonal vectors that are perpendicular to
$\vec{e}_{\rm t0}$ such that $\{ \vec{e}_{\rm t0}, \vec{e}_{\rm n0},
\vec{e}_{\rm b0}\}$ forms a right-handed basis. The displacement
components are functions of time, $t$, and arc length along the
equilibrium flux tube, $s_0$.

\subsection{Kelvin-Helmholtz instability}
\label{subsec:uniform_khi}

In the ideal case without friction between the flux tube and its
surroundings (or other dissipative processes), a uniform, straight flux
tube can only become unstable owing to the Kelvin-Helmholtz
instability. The linearized equations for the three components of the
displacement vector are all decoupled from each other. For a transverse
displacement in the normal direction we have \citep[see][]{Schmitt:1998}
\begin{equation}
   \mu \ddot{\xi}_{\rm n} + 2 u_0\dot{\xi}^{\,\prime}_{\rm n}
   + (u_0^2 - u_{\rm A}^2) {\xi}^{\prime\prime}_{\rm n} = 0\,,
   \label{eq:khi1}
\end{equation}
where $u_{\rm A} = B_0/(4\pi\rho_{\rm i0})^{1/2}$ is the Alfv{\'e}n
velocity. Dots indicate (partial) time derivatives and primes denote
derivatives with respect to arc length, $s_0$.  The factor $\mu = 1 +
(\rho_{\rm e0}/\rho_{\rm i0})$ takes into account the co-acceleration of
the surrounding medium as a consequence of a transverse acceleration of
the flux tube, leading to {\em enhanced inertia}
\citep[e.g.,][]{Spruit:1981, Moreno-Insertis:etal:1996,
Achterberg:1996a}.  Here, $\rho_{\rm e0}$ and $\rho_{\rm i0}$ are the
gas densities outside and inside the equilibrium flux tube,
respectively. The corresponding equation for the binormal component of
the displacement is  analoguous to Eq.~(\ref{eq:khi1}). The
tangential component describes (Doppler-shifted) longitudinal tube
waves, which are of no further interest in this context.

Inserting into Eq.~(\ref{eq:khi1}) an exponential {\em ansatz} of the form
$\xi_{\rm n} \propto \exp i(ks_0 - \omega t)$, with real wavenumber
$k\ge 0$ and complex frequency $\omega$, leads to the dispersion relation
\begin{equation}
   \mu\omega^2 - 2 u_0 k \omega + 
    k^2 \left (u_0^2 - u_{\rm A}^2\right ) = 0 \,.
   \label{eq:khi1disp}
\end{equation}
Instability, i.e., an exponentially growing mode, corresponds to a
positive imaginary part of $\omega$. In the case $k=0$,
Eq.~(\ref{eq:khi1disp}) only has the trivial solution, while for $k\neq
0$ we have instability if the flow velocity surpasses a critical
velocity, viz.
\begin{equation}
   u_0 > \left( {\mu\over \mu -1}\right)^{1/2} u_{\rm A}\,.
   \label{eq:khi1crit}
\end{equation}
Note that a finite value for the critical velocity requires that $\mu >
1$, i.e., that the acceleration of the surrounding medium is taken into
account. In order to identify the character of this instability, we
compare the condition derived here with the result for a `thick' flux
tube obtained by \citet{Kolesnikov:etal:2004} for incompressible
perturbations. In the thin-tube limit, $k R \ll 1$, where $R$ is the
tube radius, the dispersion relation [their Eq.~(27)] becomes identical
to our Eq.~(\ref{eq:khi1disp}), if we take into account the sign flip in
their definition of $\omega$. Consequently, our analysis provides the
thin-tube limit for the criterion for the Kelvin-Helmholtz
instability. Furthermore, \citet{Kolesnikov:etal:2004} have shown that
the critical velocity in the case of a thick tube becomes at most by
about 10\% higher than the value determined from
Eq.~(\ref{eq:khi1crit}), which therefore can be considered as a
reasonable estimate (lower limit) for all flux tubes. Similar criteria
have been given by \citet{Ryutova:1990} and also by
\citet{Achterberg:1982} and by \citet{Osin:etal:1999}, but the
connection to the Kelvin-Helmholtz instability was not made in the
latter two papers.

We have verified further (see Appendix~\ref{appendix}) that the result
does not depend on the chosen frame of reference, i.e., whether one
considers a flow within the flux tube with the exterior at rest or a
(uniform) external flow along a static flux tube. This is obvious for
the `thick' flux tube treated by \citet{Kolesnikov:etal:2004}, where the
perturbation of the exterior is consistently taken into account, in
contrast to the thin-tube case, where the effect of the exterior is
solely described by the enhanced inertia term.

\subsection{Friction-induced instability}
\label{subsec:uniform_friction}

Up to here, we have assumed the ideal case of a dissipationless
plasma. Including dissipative processes can lead to qualitatively
different behaviour. For magnetic flux tubes in stellar
convection zones, the drag resulting from a motion of the tube
perpendicular to its axis relative to the surrounding medium is an
important dissipative process. For flows with large Reynolds number, the
drag force is roughly proportional to the square of the relative
perpendicular velocity \citep[e.g.,][\S 5.11]{Parker:1975,
Batchelor:1967}. In the case of a finite perpendicular velocity in the
undisturbed flux tube equilibrium, a linear stability analysis has been
carried out in the first paper of this series
\citep{Schuessler:Ferriz-Mas:2007}. In the case of a static background
medium considered here, the drag force is of second order in the
perturbations, so that its effect cannot be studied by way of a linear
stability analysis. We therefore assume a Stokes-type friction, which is
linear in the perpendicular velocity. It turns out that the resulting
stability criteria are independent of the value of the proportionality
factor (provided that it differs from zero) and we have reason to assume
that the criteria are, in fact, independent of the nature of the
dissipation process altogether. For instance, numerical thin-tube
simulations including a quadratic drag force confirm the criteria
derived under the assumption of a Stokes-type friction (see Holzwarth et
al. in preparation, paper III of this series).  We write the Stokes-type
friction acceleration in the form
\begin{equation}
{\vec a}_{\rm S} = -\alpha\, ({\vec u}_{\rm rel}\cdot \vec{e}_{\rm n})
                   \,\vec{e}_{\rm n0} 
\label{eq:stokes}
\end{equation}
where $\alpha\ge 0$ is a constant parameter and
\beq
{\vec u}_{\rm rel} = \vec{u}_{\rm i} - \vec{u}_{\rm e}
\label{eq:urel}
\eeq
is the relative velocity, i.e., the difference between the perturbed
internal velocity, $\vec{u}_{\rm i}$, and the external velocity, $\vec{u}_{\rm
e}$. Here we have $\vec{u}_{\rm e}=\vec{0}$ and thus obtain the
linearized expression
\beqa
\vec{u}_{\rm rel} &=& u_0\vec{e}_{\rm t0} + \dot{\vec{\xi}}
            + u_0\vec{\xi}^{\,\prime} \\
        &=& (u_0 +  \dot{\xi}_{\rm t} 
             + u_0\xi_{\rm t}^{\,\prime}) \vec{e}_{\rm t0} 
             + ( \dot{\xi}_{\rm n}
             + u_0 \xi_{\rm n}^{\,\prime}) \vec{e}_{\rm n0}\,,
\label{eq:vel_perturb}
\eeqa
where
\begin{equation}
\vec{e}_{\rm n} = \vec{e}_{\rm n0}
-   \vec{e}_{\rm t0} \xi_{\rm n}^{\,\prime}
\label{eq:enorm_perturb}
\end{equation}
is the perturbed unit normal vector. The linearized part of 
${\vec a}_{\rm S}$ provides an additional term of the form $\alpha
\dot{\xi}_{\rm n}$ in Eq.~(\ref{eq:khi1}), leading to a modification of
the dispersion relation given by Eq.~(\ref{eq:khi1disp}):
\begin{equation}
   \mu\omega^2 - 2 u_0 k \omega + 
    k^2 \left (u_0^2 - u_{\rm A}^2\right ) + i\alpha\omega = 0 \,.
\label{eq:fricdisp}
\end{equation}
The resulting condition for instability (positive imaginary part of
$\omega$) is simply given by 
\begin{equation}
   u_0 >  u_{\rm A} 
\label{eq:friccrit}
\end{equation}
for $k\neq 0$. Comparison with Eq.~(\ref{eq:khi1crit}) shows that
instability in the case with friction sets in for a lower critical
velocity than for the Kelvin-Helmholtz instability, $u_{\rm KHI} =
\sqrt{\mu/(\mu-1)}u_{\rm A} > u_{\rm A}$.  This suggests that the effect of
friction leads to a new kind of instability, which does not exist in the
dissipationless case, although the condition for instability given by
Eq.~(\ref{eq:friccrit}) is independent of the actual value of the friction
parameter, as long as $\alpha$ does not vanish. In the limit $\alpha\to
0$, the growth rate of the friction-induced instability in the velocity
range $u_{\rm A} < u_0 < u_{\rm KHI}$ goes to zero, while for $u_0
>u_{\rm KHI}$ we have Kelvin-Helmholtz instability. The criterion
Eq.~(\ref{eq:friccrit}) is also independent of the enhanced-inertia
parameter, $\mu$, which indicates that the instability mechanism is not
closely related to the perturbation of the external medium, such as in
the case of the Kelvin-Helmholtz instability.

In order to elucidate the nature of the friction-induced instability, we
consider the phase velocity, $u_{\rm ph}={\rm Re}(\omega/k)$, of the
unstable mode.  It is readily shown that this mode has $u_{\rm ph}\ge 0$
for $u_0^2\ge u_{\rm A}^2$, so that the unstable mode is {\em prograde},
i.e., it propagates in the direction of the flow. For flow velocities
below the instability limit, the direction of propagation of this mode
is {\em retrograde}, against the flow. Modes whose direction of
propagation becomes reversed by the flow are sometimes called {\em
backward waves} \citep[e.g.,][]{Nakariakov:Roberts:1995,
Joarder:etal:1997}. The reversal of the propagation direction at the
instability threshold means that both real and imaginary parts of
$\omega$ change sign when $u_0 - u_{\rm A}$ goes through zero. We see
from Eq.~(\ref{eq:fricdisp}) that this corresponds also to a sign change
of the friction-related term, $i \alpha \omega$, in the dispersion
relation: the effect of friction changes from wave damping to wave
excitation.

Backward waves can also be {\em negative energy waves}
\citep[e.g.,][]{Cairns:1979,Lashmore-Davies:2005}, which owe their name
to the fact that they can grow in amplitude when the total kinetic
energy of the system (energy of the flow plus that of the wave) is
decreased (e.g., by some dissipative process). Following
\citet{Cairns:1979}, we call a given mode a negative energy wave if
the relation 
\begin{equation}
  C \equiv \omega {\partial D\over \partial \omega} < 0 
\label{eq:new}
\end{equation}
holds, where $D$ represents the left-hand side of the dispersion
relation in the non-dissipative case, here given by
Eq.~(\ref{eq:khi1disp}). In general, $D$ has to be multiplied by a
factor $\pm 1$ such that $C>0$ in the case of a system without flows
($u_0 = 0$). In our case, the condition for negative energy waves becomes
\begin{equation}
  \omega \left( \omega - {u_0 k\over \mu} \right) < 0 \,.
\label{eq:newcond}
\end{equation}
Considering the solutions of Eq.~(\ref{eq:khi1disp}),
\begin{equation}
  \omega_\pm = {u_0 k\over \mu} \pm {u_0 k\over \mu}
     \left ( 1 - \mu + {\mu u_{\rm A}^2\over u_0^2} \right )^{1/2}\,,
\label{eq:idealmodes}
\end{equation}
we find that $\omega_+$ (the prograde mode) never satisfies the
condition given by Eq.~(\ref{eq:newcond}) while $\omega_-$ (the retrograde
mode) in fact represents a negative energy wave provided that $u_0^2 >
u_{\rm A}^2$. Consequently, we find that the following three statements
are all equivalent in our case: (1) we have friction-induced
instability; (2) the system supports a backward wave mode; and (3) the
system has a negative energy wave. The fact that our unstable mode is a
negative energy wave suggests that not only the exact value of the
friction coefficient is irrelevant for the stability criterion but that
even the nature of the dissipation process is largely irrelevant. In
fact, it has been shown that energy loss by radiation of MHD waves
\citep{Ryutova:1988}, thermal conduction \citep{Joarder:etal:1997}, or
resonant absorption \citep{Tirry:etal:1998} can provide the necessary
dissipation for the development of the instability. On the other hand,
the fact that Eq.~(\ref{eq:new}) is not invariant with respect to
Galilei transformations raises questions concerning the general validity
of the concept of negative energy waves \citep[e.g.,][]{Walker:2000,
Andries:Goossens:2002}, so that we do not pursue this approach any
further.

\begin{figure}[ht!]
\centering
\resizebox{\hsize}{!}{\includegraphics[]{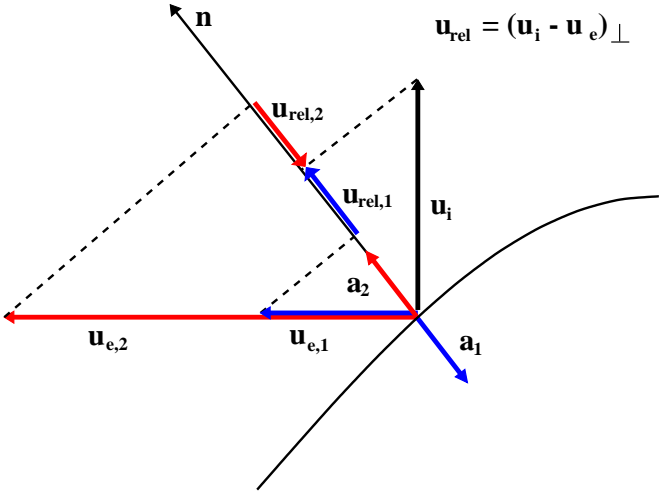}}
\caption{Schematic sketch illustrating the physical mechanism of the
friction-driven flux tube instability. A leftward propagating
transverse wave along the flux tube (indicated by the curved line) is
amplified if the speed of the leftward external flow (in the frame of
reference where the plasma in the equilibrium flux tube is at rest)
exceeds the phase velocity of the wave. See main text for a detailed
discussion. }
\label{fig:mechanism}
\end{figure}

The physical mechanism driving the friction-induced instability can be
most easily understood by considering the frame of reference in which
the plasma in the equilibrium flux tube is at rest. We then have an
external flow along the equilibrium flux tube with a velocity
$\vec{u}_{\rm e} = -u_0\vec{e}_{\rm t0}$.  
Figure~\ref{fig:mechanism} schematically shows a snapshot of a
segment of the perturbed flux tube (black curve), together with velocity
and acceleration vectors relevant for the effect of friction.  We assume
a transverse wave propagating {\em leftward}, i.e., in the direction of
the external flow. The velocity $\vec{u}_{\rm i}$ of a mass element of
the tube corresponding to that wave motion is vertically upward for the
tube segment and phase shown.  The relative velocity between the tube
and its environment that is relevant for the friction is given by the
component of the velocity difference, $\vec{u}_{\rm i} - \vec{u}_{\rm
e}$, along the instantaneous unit normal vector of the mass element,
$\vec{n}$, cf. Eqs.~(\ref{eq:stokes}) and (\ref{eq:urel}).  Depending on
the speed of the external flow, this projection can be parallel or
antiparallel to the normal vector. We consider two cases: 1) for a
sufficiently slow flow ($\vec{u}_{\rm e,1}$, blue vector arrows in
Fig.~\ref{fig:mechanism}), the relative velocity ($\vec{u}_{\rm rel,1}$)
is {\em parallel} to the normal direction, so that the acceleration by
friction ($\vec{a}_1$) is antiparallel to the normal and thus
decelerates the upward motion of the mass element, corresponding to a
damping of the wave; 2) in the case of a sufficiently fast flow
($\vec{u}_{\rm e,2}$, red vector arrows), the relative velocity
($\vec{u}_{\rm rel,2}$) becomes {\em antiparallel} to the normal
direction, so that the corresponding friction acceleration ($\vec{a}_1$)
is now parallel to the normal and accelerates the upward motion of the
mass element, thus amplifying the wave. For a wave propagating to the
right (reverse direction of $\vec{u}_{\rm i}$), we always obtain
damping.

We can put this in quantitative terms by considering the relative
velocity, 
\beq
u_{\rm rel} = (\vec{u}_{\rm i}-\vec{u}_{\rm e})\cdot\vec{e}_{\rm n}
            = \dot{\xi}_{\rm n} - u_0\xi_{\rm n}^{\,\prime}\,.
\label{eq:urel1}
\eeq
Inserting a wave solution ${\xi}_{\rm n} \propto \sin(ks_0-\omega t)$,
we readily find
\beq
u_{\rm rel} = (u_{\rm ph} + u_0) u_{\rm ph}^{-1} \dot{\xi}_{\rm n}\,,
\label{eq:urel2}
\eeq 
where $u_{\rm ph}=\omega/k$ is the phase velocity of the wave. Now, for
$u_{\rm ph}>0$ the relative velocity has always the same sign as
$\dot{\xi}_{\rm n}$, so that the wave is damped by friction. The same is
true for $u_{\rm ph}<0$ and $u_0 < -u_{\rm ph}$, while for $u_0 >
-u_{\rm ph}$ the sign of the factor in front of $\dot{\xi}_{\rm n}$ in
Eq.~(\ref{eq:urel2}) reverses, so that the friction force excites the
wave. Figuratively, one can imagine the wave being amplified by a
`tailwind'.

In the frame of reference with the external medium at rest, the
condition $u_0 > -u_{\rm ph}$ transforms into that for the `backward
wave', $u_{\rm ph}>0$.

\section{Stratified background medium}
\label{sec:stratification}

We now consider the effect of a gravitational stratification of the
external medium, so that buoyancy-driven instability
\citep{Spruit:Ballegooijen:1982,Ferriz-Mas:Schuessler:1993,
Ferriz-Mas:Schuessler:1995} is included.  For a straight, horizontal
flux tube, mechanical equilibrium then requires neutral buoyancy, i.e.,
$\rho_{\rm i0} = \rho_{\rm e0}$, in addition to the pressure balance
condition given by Eq.~(\ref{eq:pressbal}). We define the normal direction
of the equilibrium flux tube to be along the (constant) gravitational
acceleration. The linearized evolution of the three components of the
displacement vector is then governed by the following system of
equations:
\beqa
%
%
\ddot{\xi}\dw{t} + 2 u_0 \dot{\xi}\dw{t}'
+ \left( u_0^2 - u\dw{A}^2 \right) \xi\dw{t}''
-  \frac{u\dw{A}^2}{\gamma H_p}  \xi\dw{n}' = 0\,,
\label{eq:strat_tang}
\\
%
%
\mu\ddot{\xi}\dw{n} + 2 u_0 \dot{\xi}\dw{n}'
+ \left( u_0^2 - u\dw{A}^2 \right) \xi\dw{n}''
+ \alpha \dot{\xi}\dw{n} 
+ \frac{u\dw{A}^2}{\gamma H_p} \xi\dw{t}'
+ \omega_{\rm MBV}^2 \xi\dw{n} = 0\,,
\label{eq:strat_norm}
\\
%
%
\mu\ddot{\xi}\dw{b} + 2 u_0 \dot{\xi}\dw{b}'
+ \left( u_0^2 - u\dw{A}^2 \right) \xi\dw{b}''
+ \alpha \dot{\xi}\dw{b} = 0\,.
\label{eq:strat_binorm}
\eeqa
In deriving these equations%
\footnote{Describing the effect of enhanced inertia for vertical motion
  by simply introducing the constant factor $\mu$ into
  Eq.~(\ref{eq:strat_norm}) is a dubious procedure in the case of a
  stratified medium \citep{Achterberg:1996a}. Lacking a better
  treatment, we keep this crude description. The resulting stability
  criteria are independent of $\mu$; only the growth rate of modes
  dominated by the Kelvin-Helmholtz instability of normal (vertical)
  perturbations depends critically on enhanced-inertia effects, so that
  these numbers should be considered with caution.}
\citep[see also][]{Schmitt:1998}, we have assumed an ideal gas, adiabatic
perturbations, and we have taken the limit $\beta = 8\pi p_0/B_0^2\gg
1$, which reflects the conditions in stellar interiors. This simplifies
the expressions without changing the character of the various
instabilities.  In Eqs.~(\ref{eq:strat_tang})-(\ref{eq:strat_binorm}),
$\gamma$ denotes the ratio of the specific heats, $H_p$ is the pressure
scale height, $\alpha$ the friction parameter, and $\omega_{\rm MBV}$ is
the magnetic Brunt-V{\"a}is{\"a}l{\"a} frequency
\citep[cf.][]{Moreno-Insertis:etal:1992} with
\beq
\omega_{\rm MBV}^2 = \frac{u\dw{A}^2}{2 H_p^2}\left[\frac{2}{\gamma}
  \left(\frac{1}{\gamma}- \frac{1}{2}\right) - \beta\delta\right]\,,
\label{eq:mbv}
\eeq
where $\delta=\nabla - \nabla\dw{ad}$ is the superadiabaticity of the
external stratification. For $k=\alpha=0$ and $\omega_{\rm MBV}^2>0$,
the perturbed flux tube would perform stable oscillations in the normal
direction with $\omega = \omega_{\rm MBV}$. For $k=0$, $\omega_{\rm
MBV}^2>0$, and $\alpha\neq 0$, the tube behaves like a damped harmonic
oscillator.  The binormal component, Eq.~(\ref{eq:strat_binorm}), is
decoupled from the rest of the system and formally identical to the
non-stratified case treated in the preceding section, so that the same
results apply.  

For the coupled normal and tangential components we again use an
exponential {\em ansatz} of the form $\xi_{\rm t,n}(s_0,t) =
\hat{\xi}_{\rm t,n} \exp(iks_0 - i\omega t)$ and normalize all
frequencies by $k u\dw{A}$, writing $\tilde{\omega}=\omega/(k u\dw{A})$,
$\tilde{\omega}_{\rm MBV}=\omega_{\rm MBV}/(k u\dw{A})$, and
$\tilde{\alpha}=\alpha/(k u\dw{A})$. Inserting this into
Eqs.~(\ref{eq:strat_tang}) and (\ref{eq:strat_norm}) and keeping the
same symbols for the non-dimensionalized quantities, we obtain the
linear system
\beqa
\left[\tilde{\omega}^2-2M\dw{A}\tilde{\omega} 
+ \left(M\dw{A}^2-1\right)\right]
\hat{\xi}\dw{t} + iq\hat{\xi}\dw{n} = 0 
\\
\left[\mu\tilde{\omega}^2-2M\dw{A}\tilde{\omega} + \left(M\dw{A}^2-1\right)
+i\tilde{\alpha}\tilde{\omega} - \tilde{\omega}_{\rm MBV}^2\right]\hat{\xi}\dw{n} 
- iq\hat{\xi}\dw{t} = 0\,,
\label{eq:strat_coupled}
\eeqa
where $q\equiv 1/(\gamma H_p k)$ is a non-dimensional measure of the
wavelength and $M\dw{A}\equiv u_0/u\dw{A}$ is the Alfv{\'e}nic Mach number.
This leads to a dispersion relation of the form
\beq
\mu\tilde{\omega}^4+c_3\tilde{\omega}^3+c_2\tilde{\omega}^2
+ c_1\tilde{\omega}+c_0=0
\label{eq:strat_disprel}
\eeq
with
\beqa
c_3 &=& -2(\mu+1)M\dw{A} + i\tilde{\alpha} \,,
\label{eq:c3}\\
c_2 &=& (\mu+1)\left(M\dw{A}^2-1\right) + 4M\dw{A}^2
      - \tilde{\omega}_{\rm MBV}^2 - 2 i M\dw{A}\tilde{\alpha} \,,
\label{eq:c2}\\ 
c_1 &=& 2 M\dw{A} \tilde{\omega}_{\rm MBV}^2 - \left(M\dw{A}^2-1\right)
      \bigl(4M\dw{A}-i\tilde{\alpha}\bigr) \,,
\label{eq:c1}\\ 
c_0 &=& \left(M\dw{A}^2-1\right)\left( M\dw{A}^2-1 
      - \tilde{\omega}_{\rm MBV}^2\right) - q^2 \,.
\label{eq:c0}
\eeqa
\subsection{Parker instability $(\mu=1,\,\tilde{\alpha}=0)$}
\label{subsec:stratified_Parker}

In the case $\mu=1$ and $\tilde{\alpha}=0$, the transformation
$\tilde{\omega} \to \tilde{\omega} - M\dw{A}$ leads to a simple
biquadratic equation \citep{Schmitt:1998}.  The solutions show that the
flow does not affect the stability properties in this case. The
criterion for instability (positive imaginary part of $\tilde{\omega}$)
is given by
\beq
\tilde{\omega}_{\rm MBV}^2 < q^2 - 1\,,
\label{eq:Bagoo_crit}
\eeq
which corresponds to the result of \citet{Spruit:Ballegooijen:1982}. The
condition marks the onset of the buoyancy-driven undular instability or
Parker instability. In fact, the stability criterion for the Parker
instability remains unchanged for $\mu>1$. Note that, owing to our
chosen non-dimensionalisation, the wavelength-dependence of the Parker
instability is hidden in $\tilde{\omega}_{\rm MBV} = {\omega}_{\rm MBV}/(
k u_{\rm A})$ and $q=1/(\gamma H_p k)$.

\begin{figure}
\centering
\resizebox{\hsize}{!}{\includegraphics[]{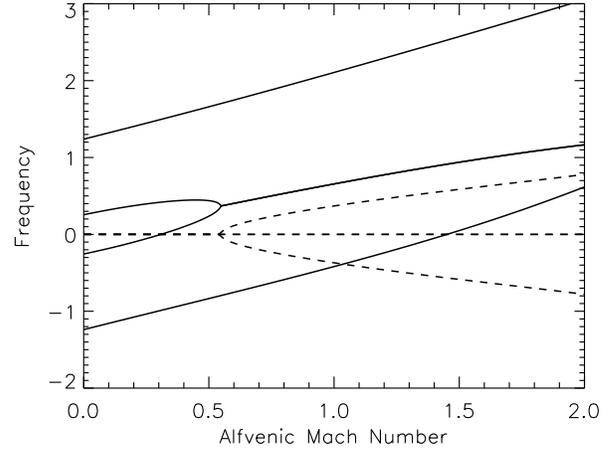}}
\caption{Kelvin-Helmholtz instability in the case with external
  stratification ($\mu=2$, $q=1$, $\tilde{\omega}_{\rm MBV}^2=0.2$)
  without friction ($\tilde{\alpha}=0$). The full and dashed lines give
  the real and imaginary parts, respectively, of the (normalized)
  frequency, $\tilde{\omega}$, for the four solutions of the dispersion
  relation, Eq.~(\ref{eq:strat_disprel}), as functions of the
  Alfv{\'e}nic Mach number of the longitudinal flow, $ M\dw{A}$.  
  Instability sets in for $ M\dw{A} \simeq 0.55$, when the real parts of
  the two (mainly) transverse modes merge and the imaginary part of
  $\tilde{\omega}$ becomes positive for one of the modes. The two modes
  whose real parts cross the zero line at $M\dw{A}\simeq 0.31$ and
  $M\dw{A}\simeq 1.45$, respectively, represent backward modes. They
  become unstable if friction is included $(\tilde{\alpha}>0)$.}
\label{fig:khi}
\end{figure}
\begin{figure}[ht!]
\resizebox{\hsize}{!}{\includegraphics[]{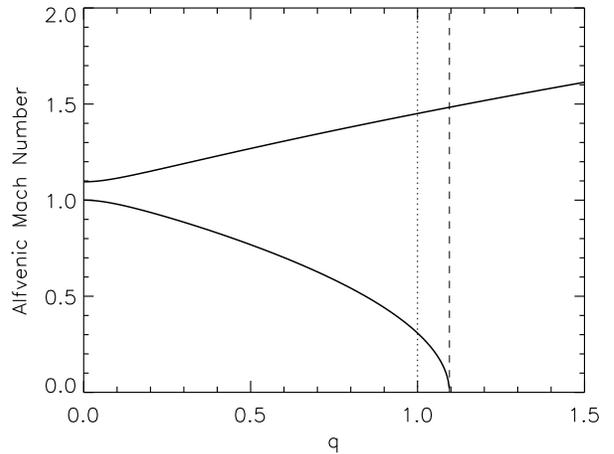}}
\caption{Critical Alfv{\'e}nic Mach numbers for the onset of the
  friction-induced instability as a function of the non-dimensionalized
  wavelength, $q$, for $\tilde{\omega}_{\rm MBV}^2=0.2$. For one of the
  two modes that can become unstable backward waves, the critical Mach
  number goes to zero as the wavelength approaches the value for the
  onset of the Parker instability, $q = (1+\tilde{\omega}_{\rm
  MBV}^2)^{1/2}$ (indicated by the dashed vertical line). The dotted
  line indicates the case $q=1$ shown in Fig.~\ref{fig:khi}. }
\label{fig:borders}
\end{figure}
\begin{figure}[ht!]
\resizebox{\hsize}{!}{\includegraphics[]{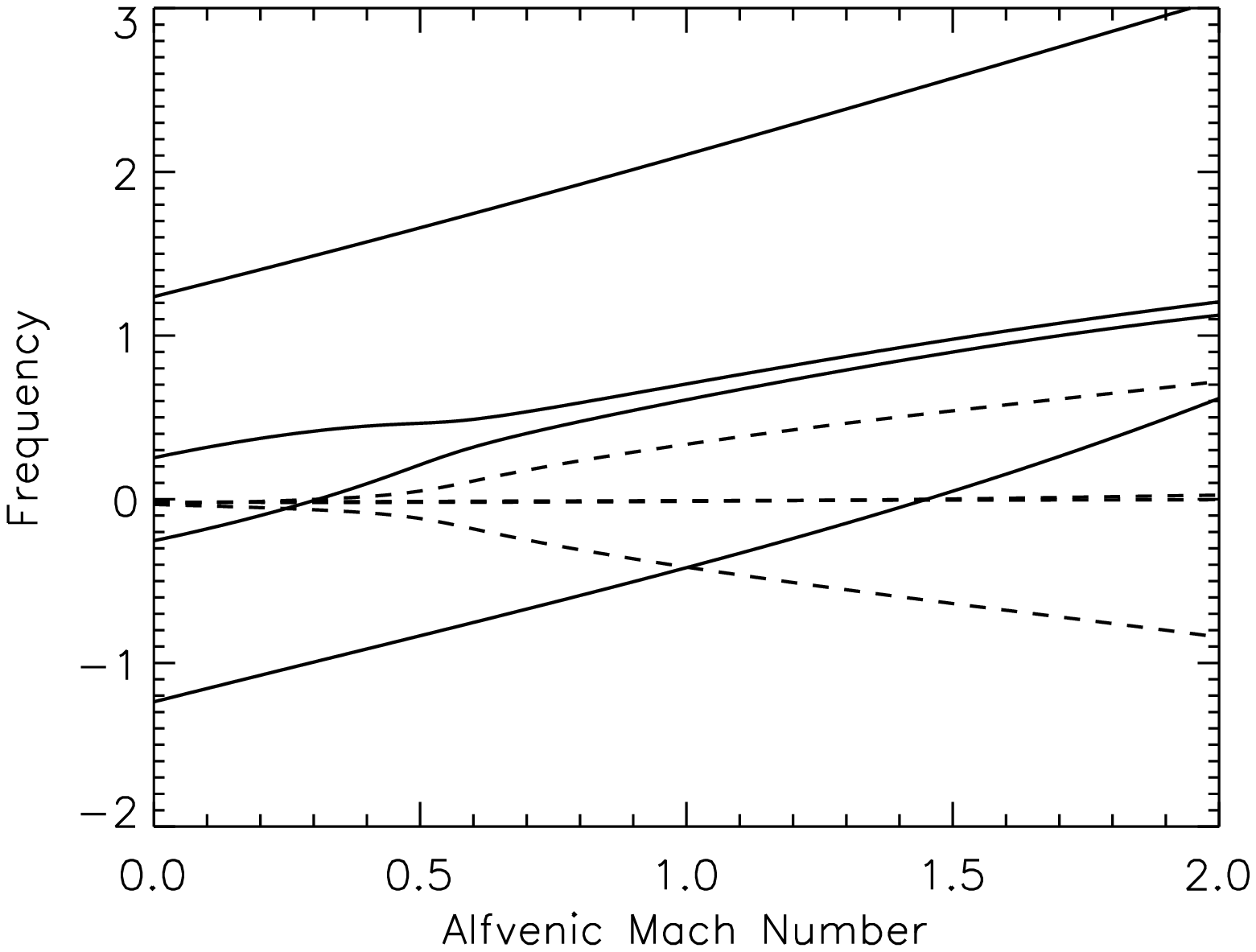}}
\resizebox{\hsize}{!}{\includegraphics[]{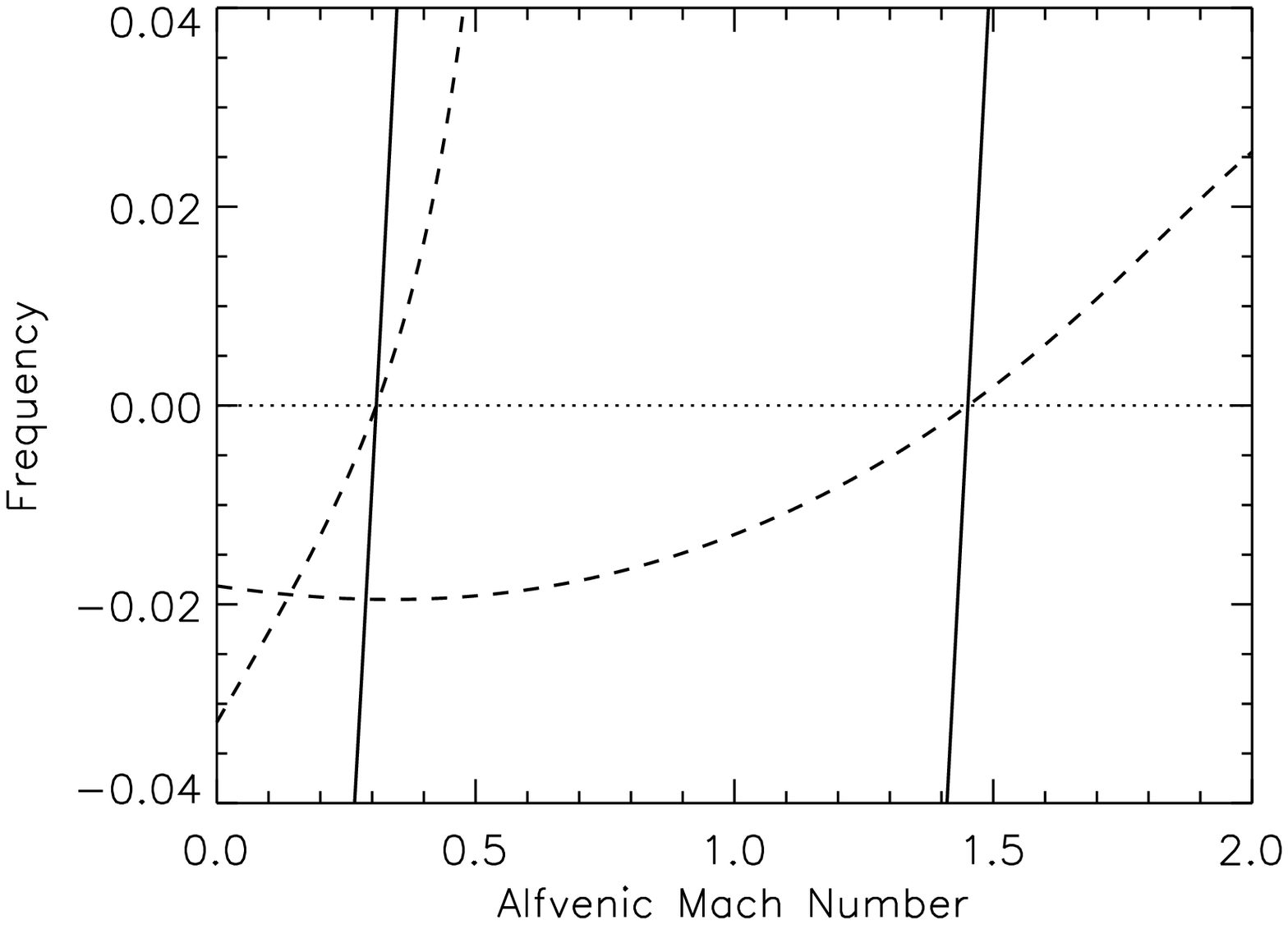}}
\caption{Kelvin-Helmholtz and friction-induced instability for the same
  case as that shown in Fig.~\ref{fig:khi}, now with
  $\tilde{\alpha}=0.2$.  The real and imaginary parts of the normalized
  frequency, $\tilde{\omega}$, are given by the full and dashed lines,
  respectively. The upper panel shows all four modes and the full range
  of frequencies, while the lower panel gives a detail of the zero
  crossing of the two backward modes.  }
\label{fig:fric}
\end{figure}

\subsection{Kelvin-Helmholtz instability $(\mu>1,\,\tilde{\alpha}=0)$}
\label{subsec:stratified_khi}
For $\mu>1$, we have the possibility of a Kelvin-Helmholtz instability
(cf. Sect.~\ref{subsec:uniform_khi}), now modified by the presence of
stratification. If the flux tube is stable with respect to the Parker
instability, i.e., if the instability condition given by
Eq.~(\ref{eq:Bagoo_crit}) is not met, Kelvin-Helmholtz instability sets
in for a flow exceeding a critical Alfv{\'e}nic Mach number, $
M\dw{A,c}$, which varies between zero for neutral Parker instability
(i.e., $\tilde{\omega}_{\rm MBV}^2 = q^2 - 1$) and infinity for
$\tilde{\omega}_{\rm MBV}^2\to\infty$. Hence, a flux tube equilibrium
approaching the stability limit of the Parker instability may become
Kelvin-Helmholtz unstable for arbitrarily low Alfv{\'e}nic Mach
number. 

The physical reason for this behavior is the fact that near the
stability limit for the Parker instability, most of the stabilizing
effect of the magnetic tension force has already been compensated for by
the stratification effects, so that a low flow velocity is sufficient to
destabilize the system. As an example, Fig.~\ref{fig:khi} shows the real
and imaginary parts of the complex frequency, $\tilde{\omega}$, as a
function of $ M\dw{A}$ for $\mu=2$, $q=1$, $\tilde{\alpha}=0$, and
$\tilde{\omega}_{\rm MBV}^2=0.2$.  In the limit $H_p\to\infty$ ($q\to
0$, ${\omega}_{\rm MBV}\to 0$, non-stratified case), the two mode pairs
represent the transverse and longitudinal tube modes
\citep{Spruit:1981}. With stratification, longitudinal and transverse
perturbations are coupled and the modes are of a mixed type.
Kelvin-Helmholtz instability sets in at $ M\dw{A} \simeq 0.55$, when the
real parts of the frequencies of one pair (corresponding to the
transverse mode in the unstratified case) merge and an unstable mode
(positive imaginary part) appears.  In the non-stratified case, the
critical Alfv{\'e}nic Mach number is significantly larger, namely equal
to $\sqrt{2}$ [cf. Eq.~(\ref{eq:khi1crit})].

\subsection{Friction-induced instability $(\tilde{\alpha}\neq 0)$}
\label{subsec:stratified_friction}

In the case $\tilde{\alpha}>0$, the effect of friction adds another
possibility for instability. Guided by the results discussed in
Sect.~\ref{subsec:uniform_friction}, we conjecture that the instability
is connected with backward waves, leading to a reversal of the sign of
the friction term in Eq.~(\ref{eq:strat_coupled}). This conjecture is
verified by direct numerical solution of the dispersion relation,
Eq.~(\ref{eq:strat_disprel}).  Fig.~\ref{fig:khi} shows that the real
part of the frequency for two modes goes from negative to positive sign,
indicating the presence of two backward waves for suffienctly high
values of $M\dw{A}$. Writing $\tilde{\omega} =
(\tilde{\omega}\dw{r},\tilde{\omega}\dw{i})$, the conditions for the
presence of a backward wave becoming marginally unstable for
$\tilde{\omega}=(0,0)$ are
\beq
\left. \frac{\mathrm{d}\tilde{\omega}\dw{r}}{\mathrm{d} M\dw{A}}
\right\vert _{\tilde{\omega}=(0,0)} > 0
\label{eq:backward_r}
\eeq
and
\beq
\left. \frac{\mathrm{d}\tilde{\omega}\dw{i}}{\mathrm{d} M\dw{A}}
\right\vert _{\tilde{\omega}=(0,0)} > 0\,.
\label{eq:backward_i}
\eeq
The dispersion relation, Eq.~\ref{eq:strat_disprel}, shows that
$\tilde{\omega}=(0,0)$ is equivalent to $c_0=0$, which entails the relation
\beq
\tilde{\omega}_{\rm MBV}^2\left( 1-M\dw{A}^2 \right) = 
q^2 - \left( 1-M\dw{A}^2 \right)^2\,.
\label{eq:c_0=0}
\eeq
Taking the derivative of Eq.~(\ref{eq:strat_disprel}) with respect to
$M\dw{A}$ and setting $\tilde{\omega}=(0,0)$ gives
\beq
c_1\left.\frac{\mathrm{d}\tilde{\omega}}{\mathrm{d} M\dw{A}}
\right\vert _{\tilde{\omega}=(0,0)} +
\frac{\mathrm{d}c_0}{\mathrm{d} M\dw{A}} = 0\,.
\label{eq:backward_cond}
\eeq
Separating this equation into its real and imaginary parts leads to
\beq
\left. \frac{\mathrm{d}\tilde{\omega}\dw{r}}{\mathrm{d} M\dw{A}}
\right\vert _{\tilde{\omega}=(0,0)} = \left(
- \frac{\mathrm{d}c_0}{\mathrm{d} M\dw{A}}
\frac{c\dw{1,r}}{c\dw{1,r}^2 + c\dw{1,i}^2}\right),
\label{eq:backward_r2}
\eeq
and
\beq
\left. \frac{\mathrm{d}\tilde{\omega}\dw{i}}{\mathrm{d} M\dw{A}}
\right\vert _{\tilde{\omega}=(0,0)} = 
- \frac{c\dw{1,i}}{c\dw{1,r}} 
\left. \frac{\mathrm{d}\tilde{\omega}\dw{r}}{\mathrm{d} M\dw{A}}
\right\vert _{\tilde{\omega}=(0,0)}
\label{eq:backward_i2}
\eeq
where $c_1=(c\dw{1,r},c\dw{1,i})$ and $c_0$ is real. Inserting the
expressions for these coefficients given in Eqs.~(\ref{eq:c1}) and
(\ref{eq:c0}) into Eq.~(\ref{eq:backward_r2}) leads to a quadratic
expression, which is positive as long as $M\dw{A}\neq 0$. Similarly, by
using Eq.~(\ref{eq:c1}) and the relation given by Eq.~(\ref{eq:c_0=0}),
one shows that $-c\dw{1,i}/c\dw{1,r}>0$, so that both conditions for
unstable backward waves given by Eqs.~(\ref{eq:backward_r}) and
(\ref{eq:backward_i}) are fulfilled. This extends our previous result
that friction-induced instability occurs in the form of unstable
backward waves to the case of a stratified medium.

The two mode pairs described by the dispersion relation
Eq.~(\ref{eq:strat_disprel}) result from the combined action of the
restoring forces of buoyancy (magnetic Brunt-V{\"a}is{\"a}l{\"a} mode)
and magnetic tension (transverse tube mode) in the system.  In both
cases, the retrograde member of the mode pair can become a backward
wave, so that we expect two modes to become friction-unstable.

In fact, as already indicated in Fig.~\ref{fig:khi}, each of the mode
pairs resulting from the dispersion relation
Eq.~(\ref{eq:strat_disprel}) exhibits a backward wave, becoming unstable
at the critical Alfv{\'e}nic Mach numbers given by the two roots of
Eq.~(\ref{eq:c_0=0}):
\beq
M\dw{A,\pm}^2 = 1 + \frac{\tilde{\omega}_{\rm MBV}^2}{2}
  \left[ 1\pm\left( 1+\frac{4q^2}
  {\tilde{\omega}_{\rm MBV}^4}\right)^{1/2}\right] .
\label{eq:backward_critMa}
\eeq
Note that, as long as they do not vanish, neither the value of
$\tilde{\alpha}$ nor that of the enhanced interia factor, $\mu$, affect
the stability criteria. It is only the growth rate of the unstable modes
that depends on the value of $\tilde{\alpha}$.

Figure~\ref{fig:borders} shows the two branches of critical Alfv{\'e}nic
Mach numbers resulting from Eq.~(\ref{eq:backward_critMa}) for
$\tilde{\omega}_{\rm MBV}^2=0.2$ as a function of the
non-dimensionalized wavelength, $q$. For $q\to 0$, i.e. wavelength
much smaller than the scale height, the lower branch corresponds to the
result for the non-stratified case (transverse mode) with $M\dw{A,-}\to
1$, while the upper branch originates from the longitudinal tube mode.
For finite wavelength, the upper branch goes to larger critical
$M\dw{A}$, while the values decrease for growing $q$ in the lower
branch. At the threshold for the onset of the Parker instability, $q =
(1+\tilde{\omega}_{\rm MBV}^2)^{1/2}$ (indicated by the dashed vertical
line in Fig.~\ref{fig:borders}), $M\dw{A,-}\to 0$.  Similar to the case
of the Kelvin-Helmholtz instability discussed in the previous
subsection, a growing part of the stabilizing effect of the curvature
force is compensated by buoyancy effects as the wavelength approaches
the critical wavelength for the Parker instability, so that lower flow
speeds suffice to trigger the friction-induced instability. The dotted
vertical line in Fig.~\ref{fig:borders} indicates the case $q=1$ shown
in Fig.~\ref{fig:khi}: the intersections of this line with the two
branches give the Mach numbers at which the corresponding two modes
become backward waves (i.e., where the real part of $\tilde{\omega}$
changes its sign).

Fig.~\ref{fig:fric} shows the mode frequencies as a function of
$M\dw{A}$ for $\mu=2$, $q=1$, $\tilde{\omega}_{\rm MBV}^2=0.2$ (same as
for the case shown in Fig.~\ref{fig:khi}), but now with a finite
friction parameter: $\tilde{\alpha}=0.2$. The upper panel gives the mode
structure in the same format as Fig.~\ref{fig:khi}. The presence of
friction removes the symmetry of the imaginary parts of $\tilde{\omega}$
(the growth rates) and also prohibits the mode merging at $M\dw{A}\simeq
0.55$ at the onset of the Kelvin-Helmholtz instability for
$\tilde{\alpha}=0$. As shown in detail in the lower panel of
Fig.~\ref{fig:fric}, the first backward mode becomes unstable for
$M\dw{A}\simeq 0.31$ and its the growth rate shows a significant upturn
for $M\dw{A} \ga 0.55$, at which value the Kelvin-Helmholtz instability
set in for $\tilde{\alpha}=0$. The second backward mode becomes unstable
at $M\dw{A}\simeq 1.45$, but the system is by then already dominated by
the first unstable mode. We note that, like in the
unstratified case, the friction-induced instability sets in earlier
(i.e., for lower values of $M\dw{A}$) then the Kelvin-Helmholtz
instability.

\section{Conclusions}
\label{sec:conclusion}

In the framework of the approximation of thin flux tubes, we have
provided a unified linear treatment of the Kelvin-Helmholtz instability,
the dissipative (friction-induced) instability, and the undulatory
(Parker-type) instability of a magnetic flux tube with a field-aligned
flow in a gravitationally stratified medium. Including dissipation
effects by a Stokes-like friction term leads to overstability of
transverse waves for flow velocities below the threshold of the
Kelvin-Helmholtz instability. In the case of a stratified medium, the
critical flow velocity can become arbitrarily low near the threshold
for Parker instability.  In other words, for a given flow velocity, there
is always a range of parameters (e.g., perturbation wavelength) that
already give friction-induced instability before the Parker-type
instability sets in.

The existence of friction-induced instability (overstability of
transverse waves) potentially has important consequences for magnetic
flux storage in stellar convection zones. Toroidal magnetic flux tubes
require faster internal rotation (equivalent to a longitudinal flow)
in order to keep force equilibrium
\citep{Moreno-Insertis:etal:1992}. Although this flow is always
sub-Alfv{\'e}nic, we have seen that it still can lead to instability in
a stratified medium. This will be studied in detail in the third paper
of this series.

To include the drag-induced dissipation in our linear analysis,
we have replaced the aerodynamic drag term (quadratic in the transverse
velocity) commonly used in treatments of thin flux tubes by a
Stokes-like friction term. As we will show by numerical simulations in
paper III, this does not affect the stability criteria for the
friction-induced instability derived here. In fact, the precise nature
of the dissipation process is not crucial for the onset of the overstability
associated with backward or negative-energy waves
\citep[cf.,][]{Joarder:etal:1997, Tirry:etal:1998}.

\appendix
\section{Kelvin-Helmholtz instability for external flow}
\label{appendix}
The results of the stability analysis have to be independent of Galilean
transformations of the frame of reference. Since, in our case, the
effect of the external medium on the flux tube is described solely by
the enhanced inertia for transverse acceleration (apart from the
pressure balance condition), it is not completely straightforward to
show that the result is Galilei invariant. Therefore, we give the
derivation in what follows.

Assume a static straight equilibrium flux tube embedded in an external
flow along the tube with $\vec{u}_{\rm e} = \varv_0 \vec{e}_{\rm
t0}$. Consider a displacement, $\xi_{\rm n}(s_0,t)$, of a Langrangean
mass element in the tube with normal acceleration $\ddot\xi_{\rm
n}(s_0,t)$. The corresponding acceleration of a mass element outside the
flux tube is given by
\beqa
\frac{\rm d}{{\rm d}t}
\left[ \frac{{\rm d}\xi_{\rm n}}{{\rm d}t}\right] &=&
\frac{\rm d}{{\rm d}t} 
\left[\dot{\xi}_{\rm n} + \varv_0 \xi_{\rm n}^\prime\right] \nonumber \\
&=& \ddot\xi_{\rm n} + 2\varv_0 \xi_{\rm n}^\prime 
+ \varv_0^2\xi_{\rm n}^{\prime\prime} \,,
\label{eq:accel}
\eeqa
where
\beq
\frac{\rm d}{{\rm d}t} = \frac{\partial}{{\partial}t} +
  \varv_0 \frac{\partial}{{\partial}s_0}\,.
\eeq
Dots indicate partial time derivatives and primes indicate partial
derivatives with respect to arc length along the undisturbed tube. By
using Eq.~(\ref{eq:accel}) and applying ``actio = reactio'', we find for
the normal component of the linearized equation of motion
\beq
\rho_{\rm i0}\ddot\xi_{\rm n} 
- \rho_{\rm i0}u_{\rm A}^2\xi_{\rm n}^{\prime\prime} 
= -\rho_{\rm e0}\left(\ddot\xi_{\rm n} + 2\varv_0 \xi_{\rm n}^\prime 
+ \varv_0^2\xi_{\rm n}^{\prime\prime}\right).
\eeq
Introducing $\mu=1+\rho_{\rm e0}/\rho_{\rm i0}$ and the {\em ansatz}
$\xi_{\rm n}\propto \exp(iks_0-i\hat\omega t)$ into this equation, we obtain
\beq
\mu\hat\omega^2 + (\mu-1)
   \left(k^2 \varv_0^2-2\varv_0 k\hat\omega\right) 
   - k^2 u_{\rm A}^2 = 0\,.
\label{eq:disprel_ext}
\eeq
We now carry out a Galilean transformation into the rest frame of the
external medium. In this frame of reference, we have an internal
longitudinal flow with velocity $u_0 \vec{e}_{\rm t0} = -\varv_0
\vec{e}_{\rm t0}$. Inserting $u_0 = -\varv_0$ and the transformation
$\omega = \hat\omega - \varv_0 k = \hat\omega + u_0 k$ into
Eq.~(\ref{eq:disprel_ext}), we find
\beq
  \mu\omega^2 - 2 u_0 k \omega + 
    k^2 \left(u_0^2 - u_{\rm A}^2\right)=0\,,
\eeq
which is identical to Eq.~(\ref{eq:khi1disp}).

\bibliography{7269.bbl}

\end{document}